\documentstyle[prl,aps]{revtex}
\input BoxedEPS.tex
\SetepsfEPSFSpecial
\HideDisplacementBoxes

\begin{document}

\twocolumn[\hsize\textwidth\columnwidth\hsize
          \csname @twocolumnfalse\endcsname
\title{Raman spectroscopic evidence for superconductivity at 645 K in 
single-walled carbon nanotubes} 
\author{Guo-meng Zhao$^{*}$} 
\address{ 
Department of Physics, Texas Center for Superconductivity and Advanced 
Materials, 
University of Houston, Houston, Texas 77204, USA}

\maketitle
\widetext
\vspace{0.3cm}

\begin{abstract}

The temperature dependent frequency shifts of the Raman active $G$-band 
have recently been measured by R.~Walter {\em et al.} for single-walled 
carbon nanotubes containing different concentrations of the magnetic 
impurity Ni:Co.  These Raman data can be quantitatively explained by  
magnetic pair-breaking effect on a superconductor with a mean-field transition 
temperature $T_{c0}$ of 645 K, in excellent agreement with independent 
electrical transport and single-particle tunneling data.  We suggest that 
such high-$T_{c}$ 
superconductivity might arise from the pairing interaction mediated 
mainly by undamped acoustic plasmons in a quasi-one-dimensional 
electronic system.

~\\
\end{abstract}
\narrowtext
]
The measurements of magnetic and electrical properties in 
multi-walled carbon  nanotube (MWNT) ropes suggest superconductivity 
above 600 K \cite{Zhao1}. This claim is buttressed by 
our recent works \cite{Zhao2,Zhao3} where we have made detailed 
analyses on a great number of existing data in literature.  We can consistently explain the temperature 
dependencies of the Hall coefficient, the magnetoresistance effect, 
the remnant magnetization, the diamagnetic susceptibility, the 
conductance, and the field dependence of the Hall voltage in terms 
of the coexistence of physically separated tubes  and  Josephson-coupled 
superconducting tubes with superconductivity above room temperature~\cite{Zhao2}.  A 
great number of the existing data for electrical transport, the 
Altshuler Aronov Spivak (AAS) and Aharonov Bohm (AB) effects, as well 
as the tunneling spectra of individual single-walled nanotubes (SWNTs) 
and MWNTs have been well explained by theories of the quantum phase 
slips (QPS) in quasi-one-dimensional superconductors \cite{Zhao3}.  
From the single-particle tunneling spectra of SWNTs, we find 
the superconducting gap  
$\Delta \simeq$ 100 meV \cite{Zhao3}.  Using a 
simple BCS relation, one obtains a mean-field transition temperature $T_{c0} \simeq$ 660 K.

We also show that, as observed in ultrathin wires of conventional 
superconductors such as PbIn and MoGe \cite{Giordano,Tinkham}, the non-zero resistance state below 
$T_{c0}$ in some individual nanotubes is similarly caused by quantum 
phase slips inherent in quasi-one-dimensional superconductors 
\cite{Zhao3},  The 
temperature dependence of the resistance in a single SWNT or MWNT is 
very similar to that in the ultrathin wires of MoGe and can be 
naturally explained by the QPS theory \cite{Zhao3}.  The resistance saturation at low temperatures observed in a single MWNT is 
a natural consequence of the QPS in quasi-one-dimensional 
superconductors \cite{Zhao3}.  Other theoretical models seem to give 
contradictory explanations to the electrical transport properties 
\cite{Sheo}.  Furthermore, the AAS effect observed in several MWNTs is 
in quantitative agreement with weak localization of Cooper pairs due 
to the large QPS \cite{Zhao3}.  In order for the observed AAS effect to 
be consistent with weak localization of single particles, one must 
assume that only the outermost layer of a MWNT is conducting 
\cite{Sheo}, in contradiction with other experiments that 
show about 14 conducting layers in a MWNT with a diameter of 
14 nm \cite{Collins}, and 27 conducting layers in a MWNT with a diameter of 
40 nm \cite{Pablo}. 

Here we analyze the data of the temperature dependent frequency shifts of 
the Raman active $G$-band in single-walled carbon nanotubes containing different 
concentrations of the magnetic impurity Ni:Co.  These data have been 
recently taken by Walter {\em et al.} at the University of North 
Carolina \cite{Walter}.  We show that these data can be quantitatively 
explained by the magnetic pair-breaking effect on a superconductor 
with $T_{c0}$$\simeq$ 645 K. The Raman data also suggest that the gap size 
is about 100 meV, in excellent agreement with independent single-particle 
tunneling data. From the deduced magnitudes 
of the gap and $T_{c0}$, we find that 2$\Delta/k_{B}T_{c0}$ $\simeq$ 3.6, in good agreement with the BCS 
prediction. 

It is known that Raman scattering has provided essential information 
about the electron-phonon coupling and the electronic pair excitation 
energy in the high-$T_{c}$ cuprate superconductors 
\cite{Krantz,Cardona,Ham}.  The anomalous temperature-dependent 
broadening of the Raman active $B_{1g}$-like mode of 90 K 
superconductors RBa$_{2}$Cu$_{3}$O$_{7-y}$ (R is a rare-earth element) 
allows one to precisely determine a superconducting gap at 2$\Delta$ = 
40.0$\pm$0.8 meV \cite{Cardona}.  Moreover, it was found that the 
threshold temperature marking the softening of the $B_{1g}$ mode with 
2$\Delta \leq \hbar\omega \leq 2.2\Delta$ coincides with $T_{c}$, and 
the mode softens further for lower temperatures.  The pronounced 
softening observed only for the $B_{1g}$ mode is due to the fact that  
the phonon energy of the $B_{1g}$ mode is very close to 2$\Delta$ and 
the mode is strongly coupled to electrons \cite{Cardona,ZZ}.  We 
emphasize that such a softening effect is observable only for 
those phonon modes with their energies very close to $2\Delta$.

In Fig.~1a, we reproduce the temperature dependence of the frequency for the 
Raman-active $B_{1g}$ mode of a 90 K superconductor 
YBa$_{2}$Cu$_{3}$O$_{7-y}$, which was reported in Ref.~\cite{Krantz}.  
It is apparent that the frequency decreases linearly with increasing 
temperature above $T_{c} \simeq$ 90 K, and that the mode starts to 
soften below $T_{c}$.  The temperature dependence of the frequency 
above $T_{c}$ is caused by thermal expansion.  The temperature 
dependence of the frequency will become more pronounced at higher 
temperatures since the magnitude of the slope $-d\ln\omega/dT$ is 
essentially proportional to the lattice heat capacity that increases 
monotonically with temperature.  The significant softening of the mode 
below $T_{c}$ occurs only if the energy of the Raman mode is very 
close to 2$\Delta$ and the electron-phonon coupling is substantial 
\cite{ZZ}, as it is the case in the 90 K superconductor 
YBa$_{2}$Cu$_{3}$O$_{7-y}$ \cite{Krantz,Cardona,Ham}.  In order to see 
more clearly the softening of the mode, we show in Fig.~1b the 
difference of the measured frequency and the linearly fitted curve 
above $T_{c}$.  It is clear that the softening starts at $T_{c}$ and 
the frequency of the mode decreases by about 9 cm$^{-1}$ at 5 K.
\begin{figure}[htb]
\input{epsf}
\epsfxsize 7cm
\centerline{\epsfbox{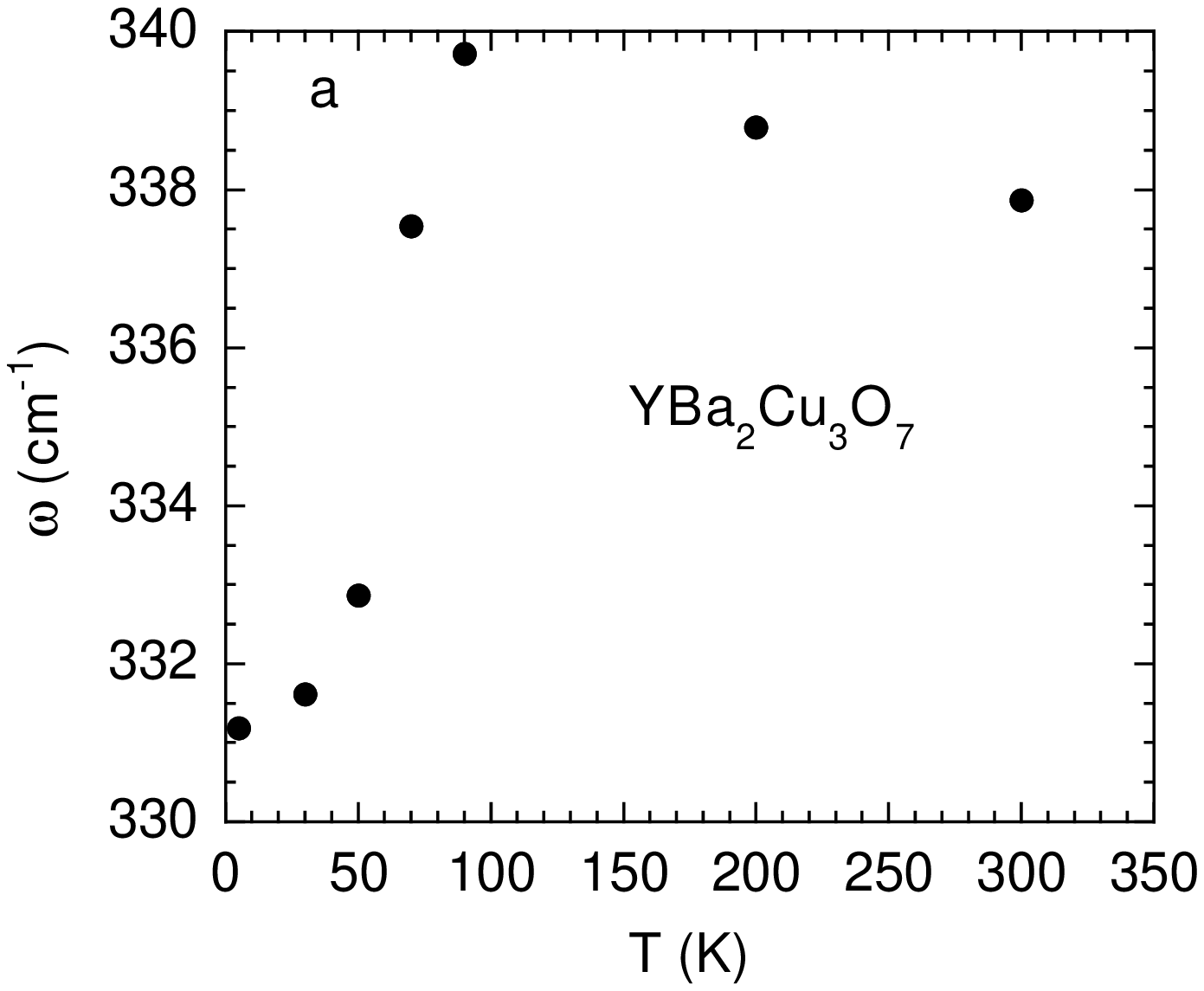}}
\input{epsf}
\epsfxsize 7cm
\centerline{\epsfbox{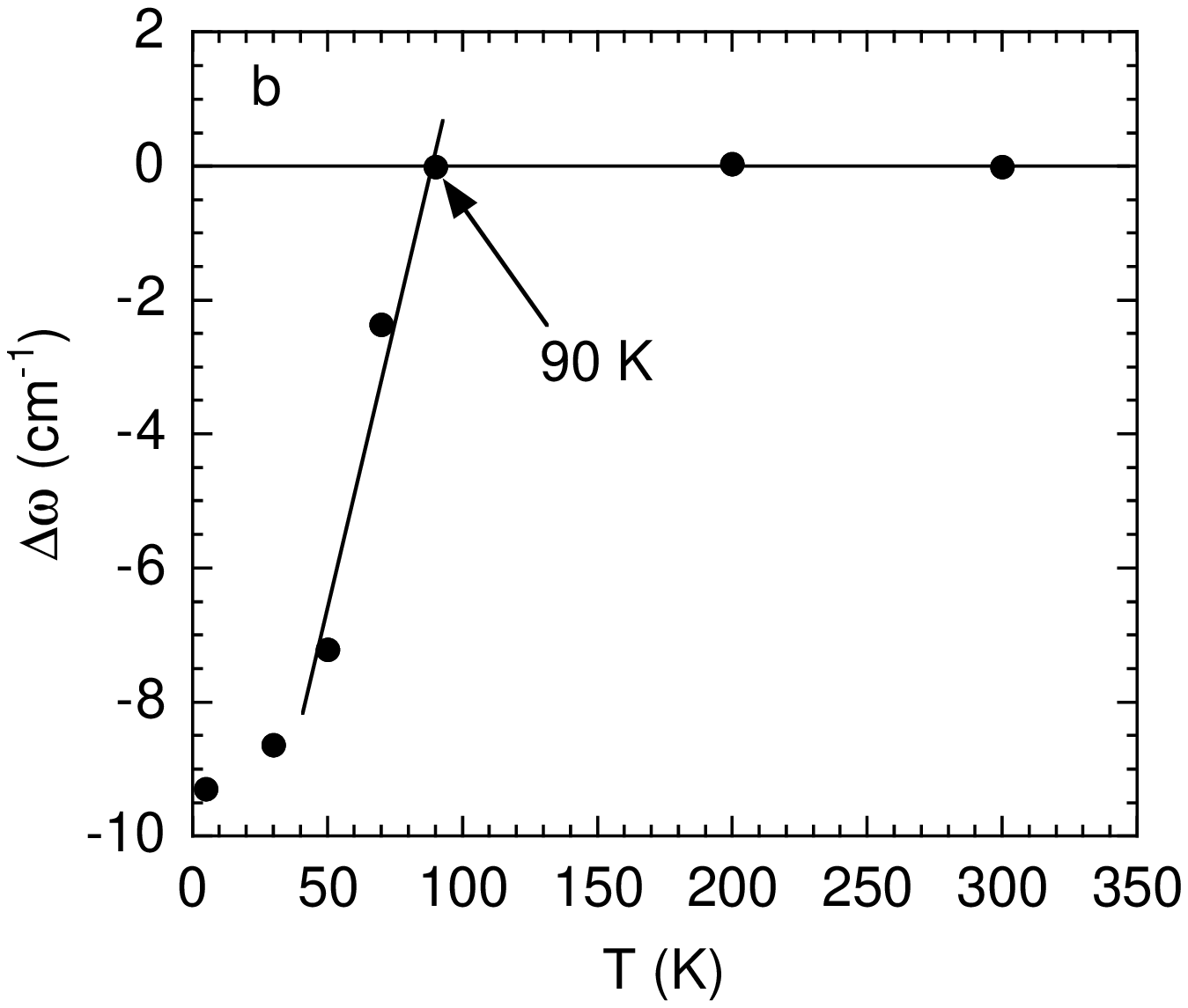}}
	\vspace{0.6cm}
	\caption [~]{a) Temperature dependence of the frequency for the 
Raman-active $B_{1g}$ mode of a 90 K superconductor 
YBa$_{2}$Cu$_{3}$O$_{7-y}$. The data are extracted from 
Ref.\cite{Krantz}.  b) The difference between the 
measured frequency and the linearly fitted curve above $T_{c}$.}
	\protect\label{fig1}
\end{figure}

Fig.~2 shows the temperature dependence of the frequency for the 
Raman active $G$-band of single-walled carbon nanotubes containing different 
concentrations of the magnetic impurity Ni:Co.  The data are from R.  
Walter {\em et al.} at the University of North Carolina \cite{Walter}.  
It is remarkable that the frequency data show a clear tendency of 
softening below about 630 K in the sample with 0.2$\%$ Ni:Co impurity.  
Above 630 K, the frequency
decreases linearly with increasing temperature, which can be 
explained as due to thermal expansion.  The larger value of 
$-d\ln\omega/dT$ in the SWNTs than in YBa$_{2}$Cu$_{3}$O$_{7-y}$ 
arises from the fact that the heat capacity above 600 K for the former is much 
larger than that for the latter below 300 K.  The merging of the 
curves in Fig.~2 at high temperatures suggests that the divergence of 
the curves at low temperatures is not due to a difference in the mean 
chirality distribution of the nanotube bundle.
\begin{figure}[htb]
\input{epsf}
\epsfxsize 7cm
\centerline{\epsfbox{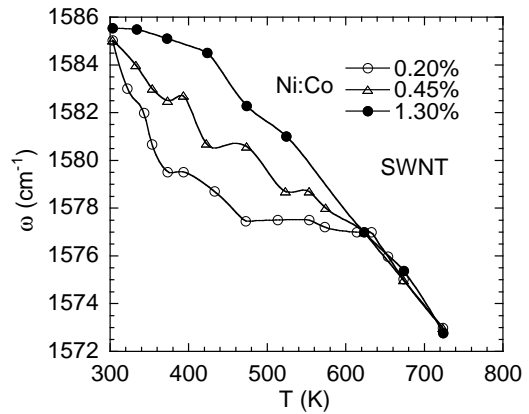}}
	\vspace{0.6cm}
	\caption [~]{Temperature dependence of the frequency for the 
Raman active $G$-band of single-walled carbon nanotubes containing different 
concentrations of the magnetic impurity Ni:Co. The curves 
are reproduced from the original plot of Ref.~\cite{Walter}.}
	\protect\label{fig2}
\end{figure}
In order to see more 
clearly the softening of the mode, we show in Fig.~3 the difference 
between the measured frequency and the linearly fitted curve above the kink 
temperatures (e.g., above 630 K for the sample containing 0.2$\%$ Ni:Co). It 
is striking that the results shown in Fig.~3 are similar to that shown in 
Fig.~1b.  This suggests that the softening of the Raman active 
$G$-band in the SWNTs may have the same microscopic origin as the 
softening of the Raman active $B_{1g}$ mode in 
YBa$_{2}$Cu$_{3}$O$_{7-y}$.  This explanation is plausible only if the 
phonon energy of the $G$-band is very close to 2$\Delta$.  Indeed, the 
phonon energy of the $G$-band is 200 meV, very close to 2$\Delta$ = 
200 meV deduced from the tunneling spectrum \cite{Zhao3}.  Therefore, it is very likely that 
the softening of the Raman active $G$-band in the SWNTs is related to 
a superconducting phase transition.

From Fig.~3, we can clearly see that the softening starts at about  632 K for the sample 
containing 0.2$\%$ Ni:Co, at about 617 K for the sample 
containing 0.45$\%$ Ni:Co, and at about 554 K for the sample 
containing 1.3$\%$ Ni:Co.  By analogy to the result shown in 
Fig.~1b, we can assign the mean-field transition temperature 
$T_{c0}$ = 632 K, 617 K, and 554 K for the samples containing 0.2$\%$, 
0.45$\%$, and 1.3$\%$ Ni:Co, respectively.
\begin{figure}[htb]
\input{epsf}
\epsfxsize 7cm
\centerline{\epsfbox{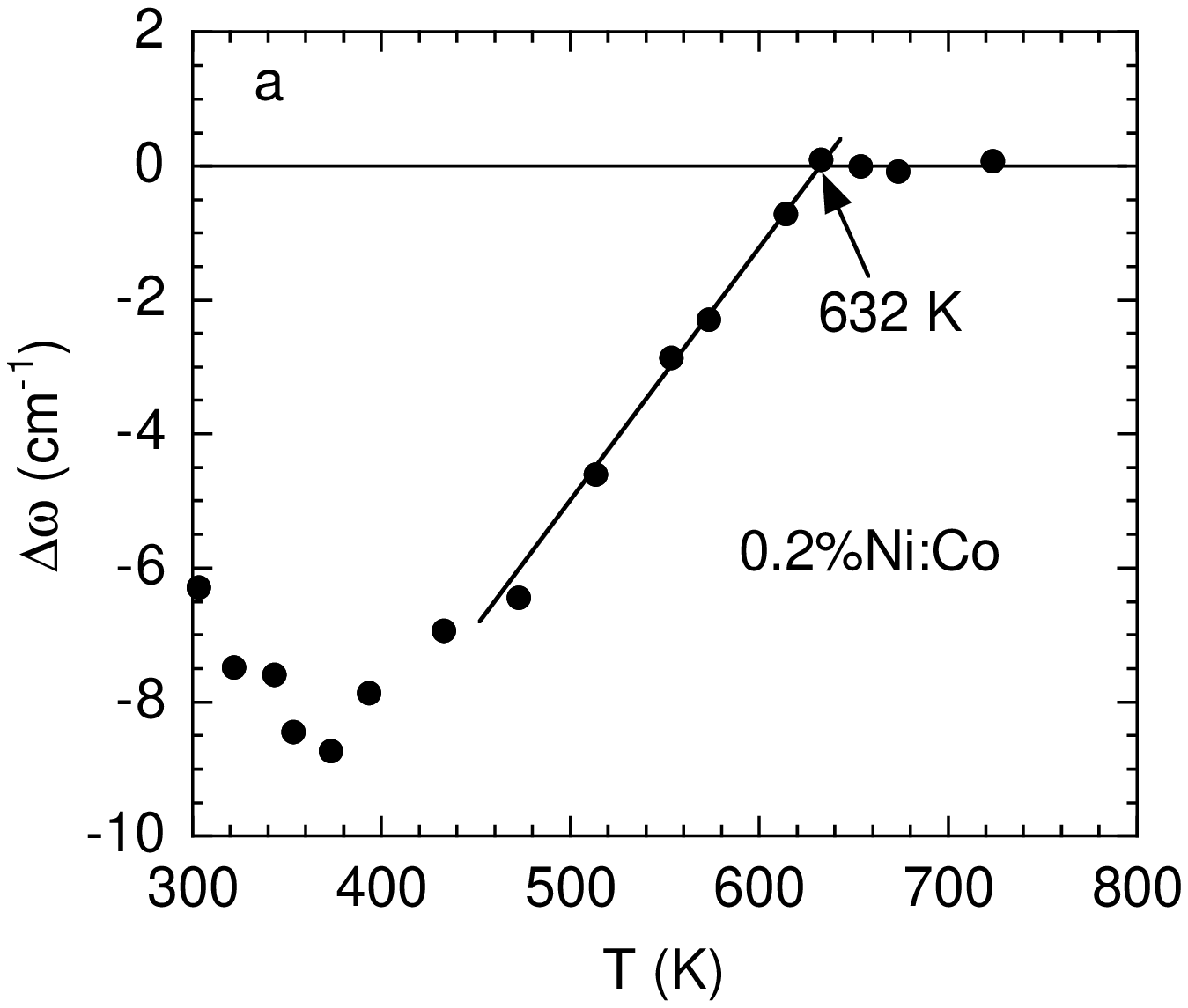}}
\input{epsf}
\epsfxsize 7cm
\centerline{\epsfbox{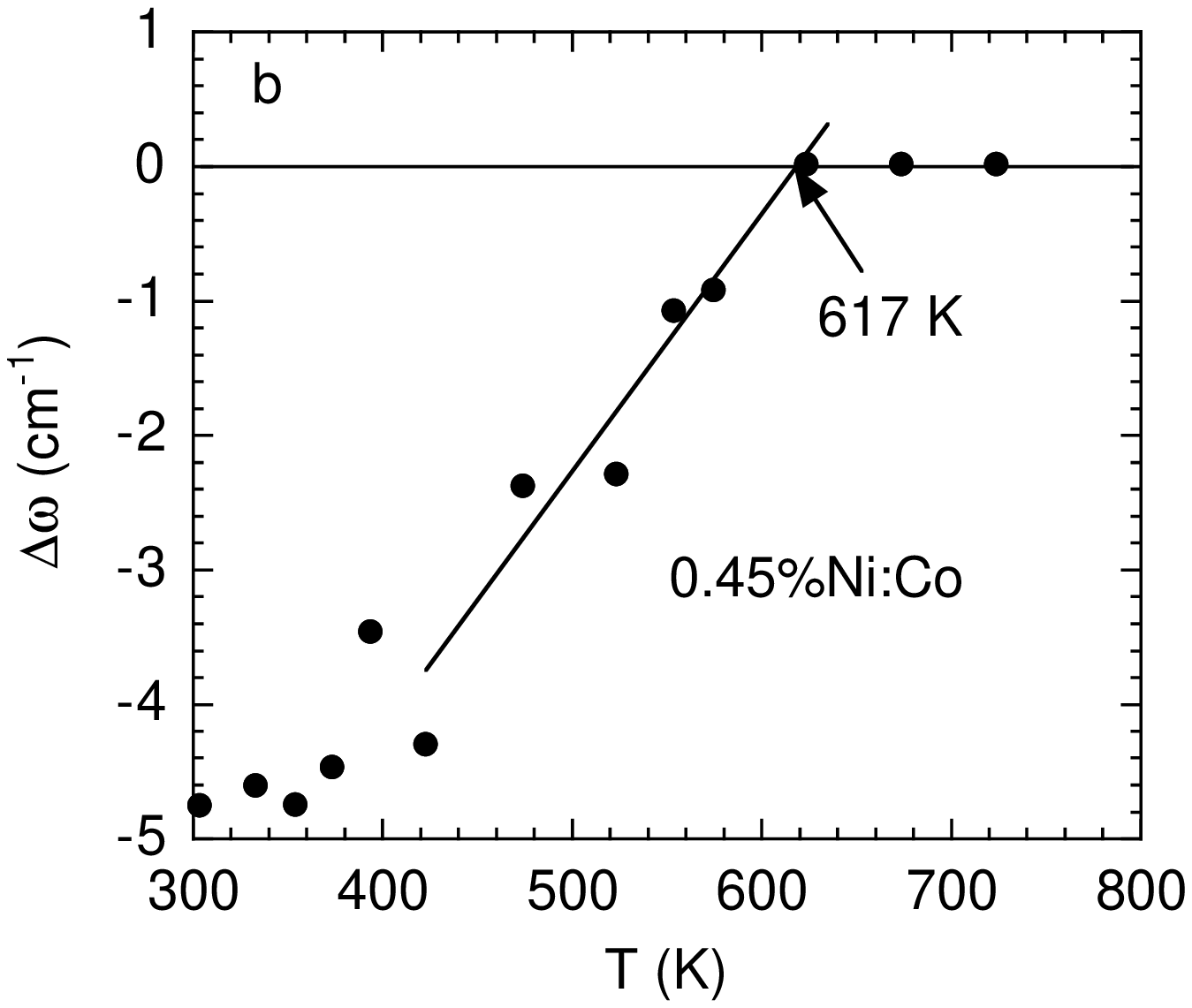}}
\input{epsf}
\epsfxsize 7cm
\centerline{\epsfbox{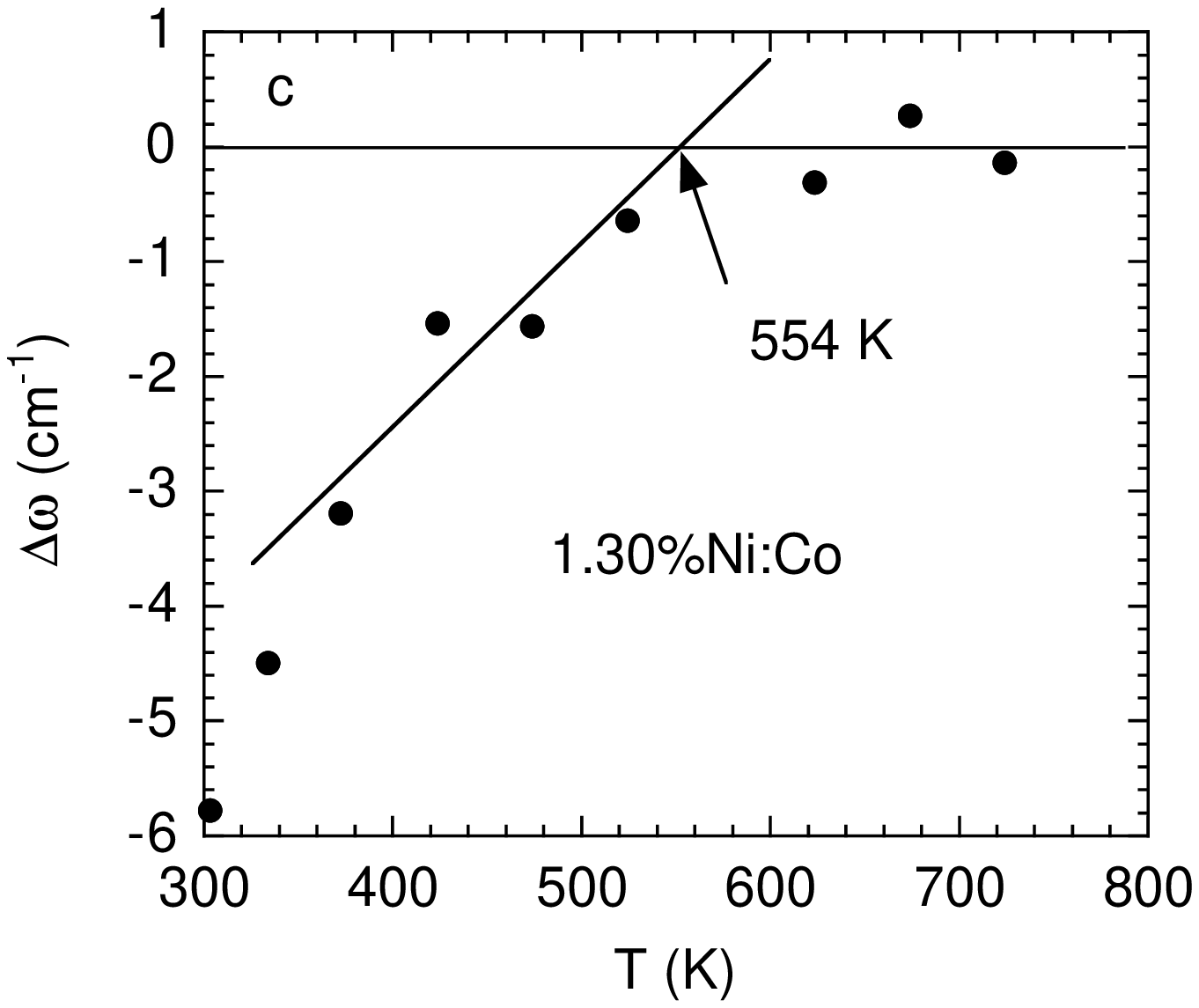}}
	\vspace{0.6cm}
	\caption [~]{The difference of the 
measured frequency and the linearly fitted curve above the kink 
temperatures (see text).}
	\protect\label{fig3}
\end{figure}

In Fig.~4, we show $T_{c0}$ as a function of the magnetic impurity 
(Ni:Co) concentration.  It is interesting that $T_{c0}$ decreases with 
increasing magnetic concentration.  The observed $T_{c0}$ dependence 
on the magnetic concentration is very similar to the theoretically 
predicted curve based on the magnetic pair-breaking effect on 
superconductivity \cite{book}.  This gives further support that the 
softening of the Raman active $G$-band in the SWNTs is related to a 
superconducting transition at around 600 K.  Extrapolating to zero 
magnetic-impurity concentration, we find $T_{c0}$ = 645 K.  Using 
$\Delta$ = 100 meV and $T_{c0}$ = 645 K, we calculate 
2$\Delta/k_{B}T_{c0}$ = 3.6, very close to that expected from the 
weak-coupling BCS theory.  It is also remarkable that the magnitude of 
the gap deduced from the Raman data is in excellent agreement with 
that inferred from a tunneling spectrum \cite{Zhao3}.

\begin{figure}[htb]
\input{epsf}
\epsfxsize 7cm
\centerline{\epsfbox{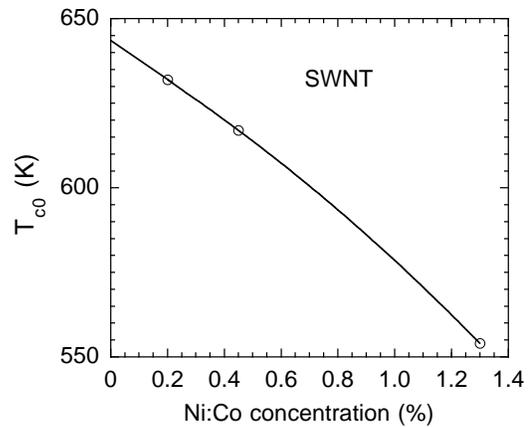}}
	\vspace{0.6cm}
	\caption [~]{Mean-field superconducting transition temperature $T_{c0}$ as a function of the magnetic impurity 
(Ni:Co) concentration in SWNTs.}
	\protect\label{fig4}
\end{figure}

It is known that the resistance of quasi-one-dimensional (quasi-1D) superconductors is finite below 
the mean-field superconducting transition temperature $T_{c0}$ due to 
quantum phase slips \cite{Zhao3,Zaikin}. In the smallest diameter 
SWNT with $d$ = 0.42 nm, the mean-field superconducting transition 
temperature $T_{c0}$ was found to be about 15 K \cite{Tang}.  The 
temperature dependence of the resistance for this 1D superconductor is 
in good agreement with the theoretical calculation ~\cite{Tang}. In 
Fig.~5a, we plot the resistance as a function of $T/T_{c0}$ 
for the smallest diameter SWNT. These data are 
extracted from Ref.~\cite{Tang}. It is apparent that the resistance  
increases more rapidly above 0.5$T_{c0}$ and flattens out towards 
$T_{c0}$. The resistance at $T_{c0}$ appears to be about four times larger 
than that at 0.5$T_{c0}$. Below 0.5$T_{c0}$, the temperature dependence of the 
resistance can be well fitted by a power law: $R(T) = R_{o} + 
AT^{\beta}$, as demonstrated in Fig.~5b. Here $R_{o}$ is contributed from 
the contact resistance and the intrinsic on-tube resistance that arises from 
the quantum phase slips.  From the fit, we find that  $\beta$ = 1.77$\pm$0.18. The theory of 
quantum phase slips in quasi-1D superconductors \cite{Zaikin} predicts that $\beta = 
2\mu -3$, where $\mu$ is a quantity that characterizes the ground state.  
The on-tube resistance at zero temperature can approach zero when $\mu > 2$, 
but is finite when $\mu < 2$.  Disorder can lead to weak 
localization of Cooper pairs and thus make $\mu < 2$  \cite{Zaikin,Zhao3}. 
\begin{figure}[htb]
\input{epsf}
\epsfxsize 7cm
\centerline{\epsfbox{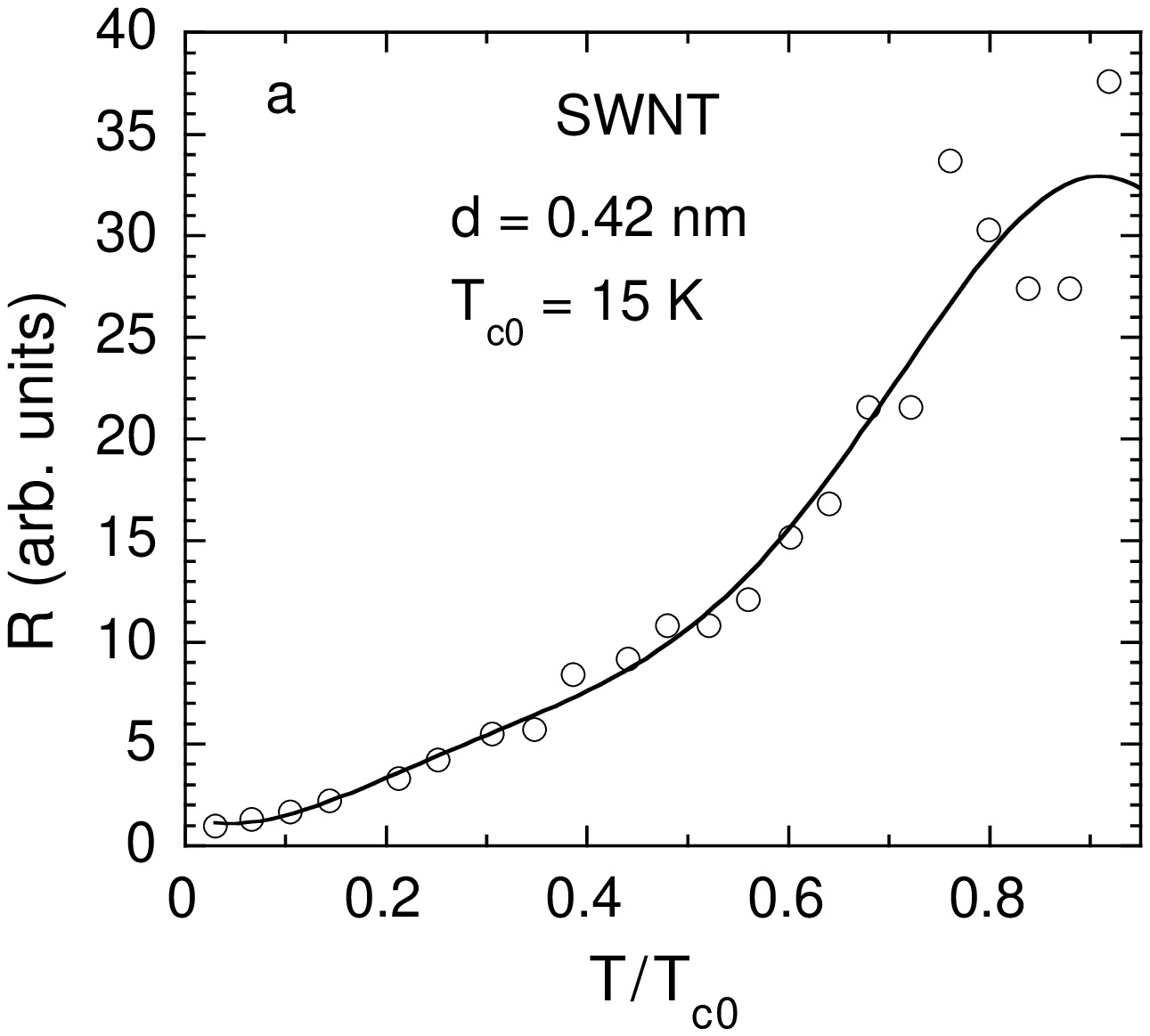}}
\input{epsf}
\epsfxsize 7cm
\centerline{\epsfbox{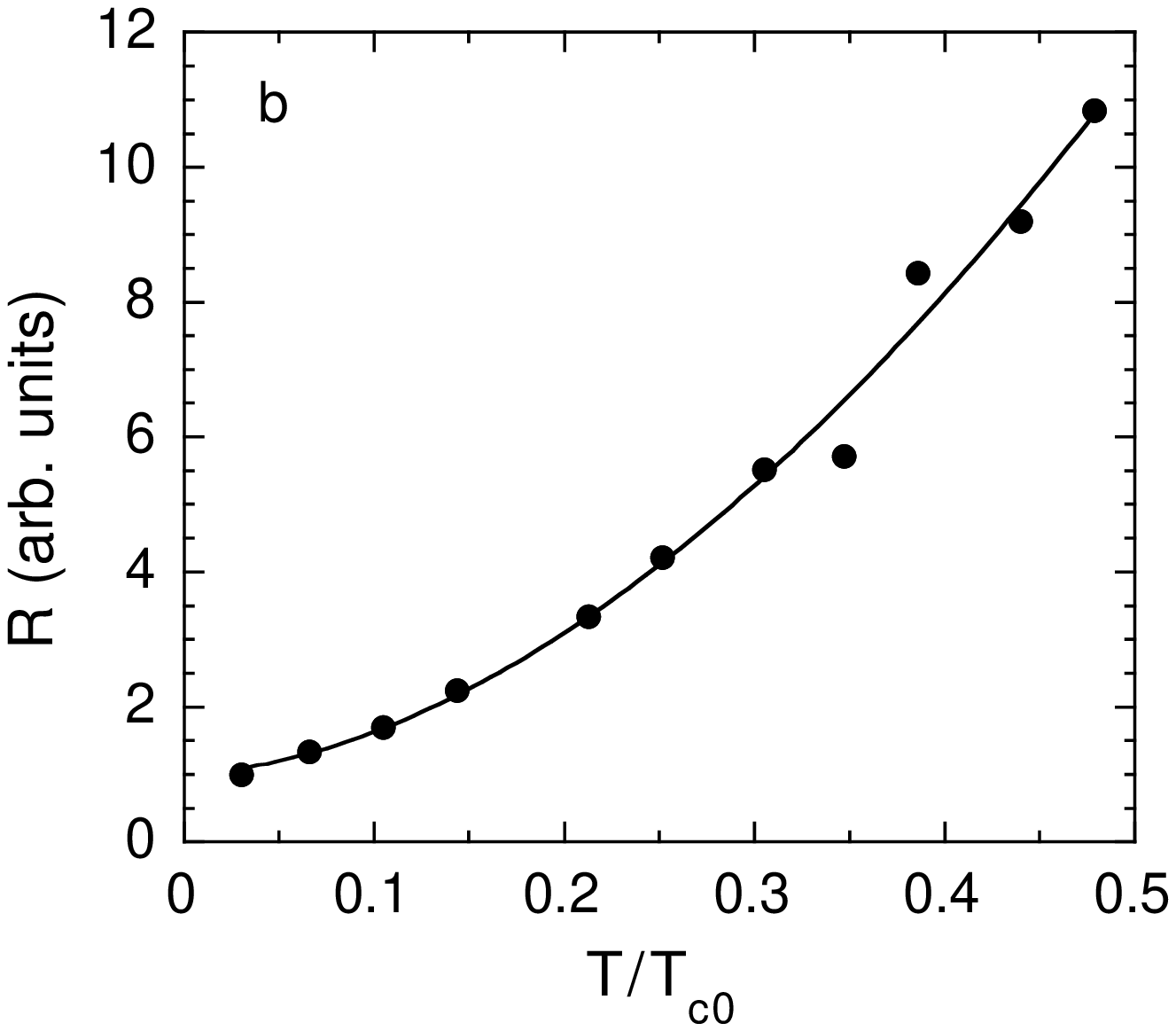}}
	\vspace{0.6cm}
	\caption [~]{a) The resistance data as a function of $T/T_{c0}$ 
for the smallest diameter SWNT with $d$ = 0.42 nm. The data are 
extracted from Ref.~\cite{Tang}. b) The temperature dependence of the 
resistance below 0.5$T_{c0}$. The data can be well fitted by $R(T) = R_{o} + 
AT^{\beta}$ with $\beta$ = 1.77$\pm$0.18. }
	\protect\label{fig5}
\end{figure}
In Fig.~6, we show the temperature dependence of the resistivity for a 
SWNT rope. These data are extracted from Ref.~\cite{Nature}. Below 
200 K, the resistivity is nearly temperature independent, which 
suggests that the measured resistance is contributed only from 
superconducting SWNTs with metallic chiralities. Since the resistance for semiconducting chirality 
tubes is larger than that for the metallic chirality tubes by several orders of 
magnitude \cite{Paul}, any current paths which include semiconducting chirality tubes are 
``shorted'' by current paths which consist of only superconducting tubes. 
Considering the fact that two thirds of the tubes have semiconducting 
chiralities, the intrinsic resistivity of the metallic chirality tubes must 
be much smaller than that shown in Fig.~6.
The contact barriers among the metallic chirality tubes may contribute to the 
resistance that increases weakly with decreasing temperature 
\cite{Baum}. The 
nearly temperature independent resistance observed below 200 K might be due to the competing contributions of the barrier 
resistance and on-tube metal-like resistance below $T_{c0}$ (due to 
quantum phase slips).   Above 200 K, the resistivity increases 
suddenly and starts to flatten out above 
550 K. Such a resistive temperature dependence is similar to that 
shown in Fig.~5a, and is consistent with quasi-1D superconductivity 
with $T_{c0}$ $\simeq$ 600 K.
\begin{figure}[htb]
\input{epsf}
\epsfxsize 7cm
\centerline{\epsfbox{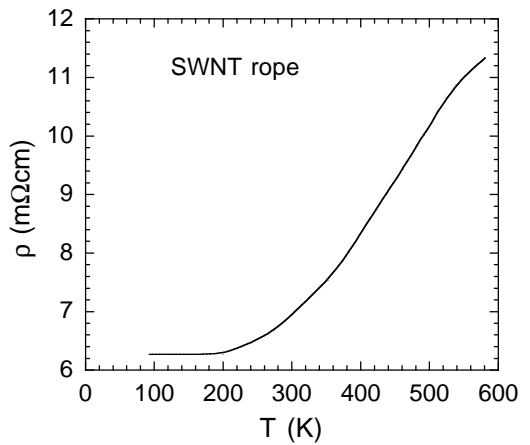}}
	\vspace{0.6cm}
	\caption [~]{Temperature dependence of the resistivity for a 
SWNT rope. The data are extracted from Ref.~\cite{Nature}.}
	\protect\label{fig6}
\end{figure}

We would like to mention that only armchair tubes 
have zero gap while other metallic chirality tubes have small 
semiconducting gaps due to a finite curvature \cite{KanePRL}.  These metallic 
chirality tubes will superconduct when the doping lever is sufficient 
to move the Fermi level away from the small gap range.  In most cases, 
the intrinsic defect-mediated doping is enough to drive the metallic 
chirality tubes to superconduct at a temperature well above room 
temperature.  If the semiconducting chirality tubes could become 
superconducting by sufficient doping, the resistivity below $T_{c0}$ 
would be still very large because the quantum phase slips are very 
significant due to a small number of transverse channels and a large 
normal-state resistivity \cite{Zaikin,Zhao3}.  The semiconducting 
chirality tubes would also act to separate and weaken the Josephson 
coupling between the superconducting tubes.

The temperature dependence of the resistance for a 
single-walled nanotube with $d$ = 1.5 nm is shown in Fig.~7. These data are extracted from 
Ref.\cite{Kong}. The distance between the two contacts is about 200 nm 
and the contacts are nearly ideal with the transmission probability of 
about 1 \cite{Kong}. It is remarkable that the temperature dependence of the 
resistance can be fitted by a power law: $R(T) = R_{o} + 
AT^{\beta}$ with $\beta$ = 1.71$\pm$0.23. The power $\beta$ for the 
1.5 nm SWNT is nearly the same as that for the 0.4 nm SWNT, which has 
been proved to be superconducting. Comparing Fig.~7 with Fig.~5, we 
could infer that $T_{c0}$ for the 1.5 nm SWNT is above 600 K. Within 
the theory of quantum phase slips in quasi-1D superconductors \cite{Zaikin}, $\beta = 
2\mu -3$,  so we have $\mu$ = 2.36$\pm$0.12  with  $\beta$ = 1.71$\pm$0.23.  
The value of $\mu$ $>$ 2 implies zero on-tube 
resistance at zero temperature from the theory \cite{Zaikin}, in good 
agreement with the experimental result \cite{Kong}.
\begin{figure}[htb]
\input{epsf}
\epsfxsize 7cm
\centerline{\epsfbox{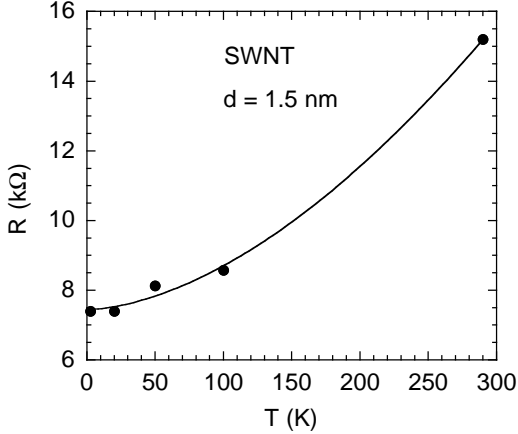}}
	\vspace{0.6cm}
	\caption [~]{Temperature dependence of the resistance for a 
single-walled nanotube with $d$ = 1.5 nm. The data are extracted from 
Fig.~1a of Ref.~\cite{Kong} at zero gate voltage where the Fermi 
level is at least 0.2 eV from the band center.}
	\protect\label{fig7}
\end{figure}

It is also interesting to note that the data shown in Fig.~6 and Fig.~7 are not compatible with  Luttinger-liquid  
behavior. Phonon backscattering in Luttinger liquid 
leads to semiconductor-like electrical transport at low temperatures and  
to metal-like electrical transport  with 0.5$<$$\beta$$<$ 1 at high 
temperatures \cite{Kane,Komnik}. This 
is in contrast with the data shown in Fig.~6 and Fig.~7, which 
suggest $\beta$ $>$ 1. Furthermore, the resistance at zero gate voltage is even temperature 
independent from 2 K to 270 K for a SWNT with a length of about 
1~$\mu$m (see the inset of Fig.~1b of 
Ref.~\cite{Kong}). This implies that $\beta$ = 0 over the wide 
temperature region of 2-270 K.  Such an unusual temperature 
dependence of the on-tube resistance cannot be explained by Luttinger 
liquid theories, but can be naturally explained by the theory of quantum 
phase slips in quasi-1D superconductors which predicts $\beta$ = 0 in 
the case of $\mu$ = 1.5.   In addition, the scanning tunnelling 
microscopy on individual undoped armchair SWNTs 
shows no pseudo-gap feature in electronic density of states at the 
Fermi level \cite{Lieber}. This experimental result is in contrast with 
the Luttinger-liquid theory 
by Kane {\em et al.} \cite{Kane} where the Luttinger parameter $g$ 
is predicted to be 0.2-0.3, but may be consistent with the Luttinger 
theory by Konik {\em et al.} \cite{Konik} where the Luttinger 
parameter is predicted to be close to 1 at any doping levels.  

On the other hand, tunneling spectra for  
SWNT bundles \cite{Bock} and individual MWNTs \cite{Bachtold2001} indicate 
that $\alpha^{end}$ $\simeq$ 2$\alpha^{bulk}$ and 
$\alpha^{bulk}$$\simeq$ 0.3 at temperatures below 100 K, where 
$\alpha^{bulk}$/$\alpha^{end}$ is the exponent of power law in 
tunneling spectrum for the electron tunneling into the bulk/end of the 
tubes.  Although the value of $\alpha^{bulk}$$\simeq$ 0.3 appears to agree 
with the prediction of the Luttinger liquid theories \cite{Kane,Komnik}, 
the value of $\alpha^{end}$ does not.  In order to satisfy the 
experimentally observed condition: $\alpha^{end}$ $\simeq$ 
2$\alpha^{bulk}$, one must assume 
that electrical transport should involve many one-dimensional 
subbands \cite{Bachtold2001}, which would make the electronic system 
quasi-3 dimensional and unfavorable to the Luttinger liquid behavior. 
Alternatively, the environmental 
Coulomb blockade theory can well explain the experimental data \cite{Bachtold2001}.

We should mention that only one group claimed \cite{PM} that both $\alpha^{bulk}$ 
and $\alpha^{end}$ (i.e., $\alpha^{end}$ $\simeq$ 2.7$\alpha^{bulk}$) 
for an individual SWNT are consistent with the Luttinger liquid model. 
There are several problems with the claim. First of all, they deduced values 
of $\alpha^{bulk}$ and $\alpha^{end}$ from the high-temperature data 
(above 120 K), while the Luttinger liquid theory \cite{Komnik} predicts that $\alpha^{end}$ 
$\simeq$ $\alpha^{bulk}$ in this high temperature regime. Second, 
the experimental data appear to indicate that $\alpha^{end}$ is 
temperature independent above 120 K, in contrast with the theoretical 
prediction \cite{Komnik}.  Third, the experimental data suggest that 
$\alpha^{bulk}$ is temperature dependent \cite{PM}, while the authors of 
Ref.~\cite{PM} approximated with a single 
exponent in the whole temperature range of 120-300 K.  If one corrects the 
Coulomb blockade contribution for the data at low temperatures as the 
authors of Ref.~\cite{Bock} did, one may find from the data of 
Ref.~\cite{PM} that $\alpha^{end}$ $\simeq$ 2$\alpha^{bulk}$ at low 
temperatures, which is actually inconsistent with the Luttinger liquid 
theory.

If the Luttinger liquid behavior is not relevant to carbon 
nanotubes, the mechanism for the non-Luttinger liquid behavior may be 
a strong nonretarded electron-phonon interaction which would lead to 
an effectively attractive interaction between two electrons. In doped 
C$_{60}$, the phonon energy is comparable with the Fermi energy, the 
electron-phonon interaction is essentially nonretarded. 
High-temperature superconductivity may arise from the strong non-retarded electron-phonon 
interaction that is large enough to overcome the direct Coulomb repulsive 
interaction.  For doped carbon nanotubes, the phonon energy of the 
$G$-band is larger than or comparable with the Fermi energy in low 
doping range so that a strong nonretarded electron-phonon interaction 
may give rise to an effectively attractive interaction between two 
electrons.  This electron-phonon interaction alone would not lead to 
high-temperature superconductivity due to a low density of states 
at the Fermi level.

Now a question arises: What is the pairing mechanism responsible for 
such high superconductivity in carbon nanotubes, and why does the 
smallest SWNT have a much lower $T_{c0}$? A theoretical calculation showed that 
superconductivity as high as 500 K can be reached through the pairing 
interaction mediated by acoustic plasmon modes in 
a quasi-one-dimensional electronic system \cite{Lee}. The calculated 
$T_{c}$ as a function of the areal 
carrier density for InSb wires of the cross sections of 50 
nm$\times$10 nm and 80 nm$\times$10 nm is reproduced in Fig.~8.  This calculation indicates that the highest $T_{c}$ occurs at a 
doping level where the first 1D subband is nearly occupied, and that 
superconductivity decreases rapidly with increasing carrier density. 
This is because an increase of the carrier density raises the Fermi 
level so that more transverse levels are involved, diminishing the 
quasi-1D  character of the system.  For a metallic single-walled nanotube 
with $d$ $>$ 1 nm, two degenerate 1D subbands are partially occupied by 
hole carriers with the carrier concentration in the order of 
10$^{19}$/cm$^{3}$. This is the most favorable condition for achieving 
high-temperature superconductivity within the plasmon-mediated 
mechanism \cite{Lee}. On the other hand, the smallest SWNT has a carrier density 
of 3.4$\times$10$^{23}$/cm$^{3}$, as estimated from the measured 
penetration depth (3.9 nm) and the effective mass of supercarriers 
(0.36 m$_{e}$) \cite{Tang}.  One can easily show that 8 
transverse subbands cross the Fermi level in the smallest SWNT, which 
makes the plasmon-mediated mechanism very ineffective. This can 
naturally explain why the $T_{c0}$ in the smallest SWNT is only 15 K. 
Interestingly, the value 
2$\Delta/k_{B}T_{c0}$ = 3.6 deduced for SWNTs is in remarkably 
good agreement with the theoretical prediction \cite{Lee}.
\begin{figure}[htb]
\input{epsf}
\epsfxsize 7cm
\centerline{\epsfbox{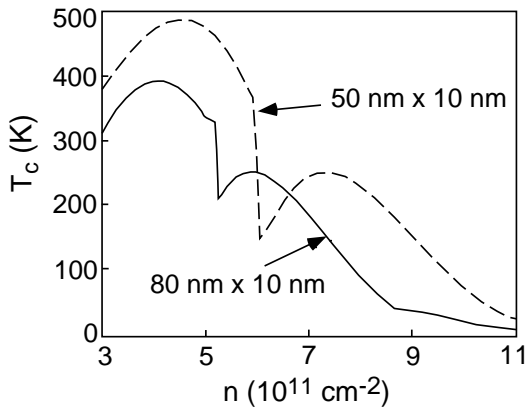}}
	\vspace{0.6cm}
	\caption [~]{The calculated $T_{c}$ as a function of the areal carrier 
density for InSb wires of the cross sections of 50 nm$\times$10 nm 
and 80 nm$\times$10 nm. The curves 
are reproduced from the original plot of Ref.~\cite{Lee}.}
	\protect\label{fig8}
\end{figure}

For multi-layer electronic 
systems such as cuprates and MWNTs, high-temperature superconductivity 
can occur due to an 
attraction of the carriers in the same conducting layer via exchange 
of virtual plasmons in neighboring layers \cite{Cui}.  Indeed, a strong 
coupling of electrons with high-energy ($\sim$ 2eV) electronic excitations in 
cuprates has been shown by well-designed optical experiments 
\cite{Little}.  For MWNTs, the dual character of the 
quasi-one-dimensional and multi-layer electronic structure could lead 
to a larger pairing interaction and a higher $T_{c0}$.  It is 
interesting that the energy gap (pairing energy) in the carbon 
nanotubes is close to that ($>$ 60 meV) \cite{gap} for deeply 
underdoped cuprates.  Analogously, we hold that these deeply 
underdoped cuprates would exhibit phase-coherent superconductivity 
above room temperature if the superfluid density could be enhanced by 
a factor of 3.

In summary, we have analyzed the data of the temperature dependent frequency 
shifts of the Raman active $G$-band in single-walled carbon nanotubes 
containing different concentrations of the magnetic impurity Ni:Co.  
The data can be 
quantitatively explained by the magnetic pair-breaking effect on 
superconductivity with a mean-field transition temperature of 645 K 
and 2$\Delta/k_{B}T_{c0}$ = 3.6.  We suggest that such high temperature superconductivity might arise from the pairing interaction mediated 
mainly by  acoustic plasmons in quasi-one-dimensional 
electronic system. ~\\
 ~\\
{\bf Acknowledgment:} I am grateful to Dr. R. Walter {\em et al.} for sending me 
their unpublished data. I thank Dr. Pieder Beeli for 
for his critical reading and comments on the manuscript. The author acknowledges financial support from the State of Texas 
through the Texas Center for Superconductivity and Advanced Materials at 
the University of Houston where some of the work was completed.
 ~\\
 ~\\  
* Correspondence should be addressed to gmzhao@uh.edu.

\end{document}